\documentclass{article}
\usepackage{spconf,amsmath,graphicx,hyperref}
 \usepackage{caption}
 \usepackage{subcaption}
 \usepackage{wrapfig}
 \usepackage{multirow}
 \usepackage{enumitem}
 \usepackage{xcolor}         
\usepackage{amsmath}
\usepackage{algorithm}
\usepackage{algpseudocode}
\usepackage{color}

\usepackage{tcolorbox}
\usepackage{url}            
\usepackage{booktabs}       
\usepackage{amsfonts}       
\usepackage{nicefrac}       

\title{Robust Workflow Generation via Adversarial Learning for Audio Deepfake Detection}
\name{Xiang Li$^1$, Pin-Yu Chen$^2$, Wenqi Wei$^1$}
\address{$^1$Department of Computer Science and Information, Fordham University, NY, USA \\
$^2$IBM Research, Yorktown Heights, NY, USA}

\begin{document}
%
\maketitle

\begin{abstract}
The rapid advancement of speech synthesis and voice conversion technologies has made audio deepfakes increasingly realistic, posing serious security risks in practical applications. While existing detection methods achieve strong performance under controlled conditions, they often fail to generalize under real-world perturbations and corruptions. In this paper, we propose \textbf{ROGUE}, a framework that dynamically constructs robust detection workflows by orchestrating multiple detection tools. ROGUE formulates workflow generation as a sequential decision-making problem and introduces a dual-agent paradigm, where a perturbation agent generates audio perturbations and a policy agent learns to select and execute detection tools under perturbed conditions. Through adversarial learning, ROGUE enables perturbation-aware tool selection, adaptive execution strategies, and improved robustness to distribution shifts. Extensive experiments across multiple datasets and real-world corruptions demonstrate that ROGUE consistently outperforms strong baselines in both robustness and generalization. Our results highlight the effectiveness of adversarially optimized workflow generation for building reliable audio deepfake detection systems in real-world deployment settings.
\end{abstract}
\section{Introduction}
\label{sec:intro}

The rapid advancement of Text-to-Speech (TTS) and Voice-Conversion (VC) using Artificial Intelligence (AI) technology has enabled highly realistic speech generation at scale~\cite{vyas2023audiobox, ye2024flashspeech}. While these technologies offer substantial benefits for applications such as content creation and accessibility, they also introduce serious security risks, including impersonation, misinformation, and social engineering attacks.

To mitigate these risks, audio deepfake detection has become critical for safeguarding digital communication. Existing detection methods range from signal processing techniques, spectral artifact analysis, and deep neural models such as self-supervised speech foundation models~\cite{wang2021investigating,zhang2024audio,li2025we}. Despite strong detection performance under controlled conditions, these detectors often degrade substantially under real-world corruptions such as noise, compression, reverberation, and codec transformations~\cite{li2025measuring}. Moreover, no single detector performs consistently well across all conditions, motivating the development of multi-detector systems that aim to leverage complementary strengths.

In parallel, Large Language Model (LLM) agents have demonstrated strong capabilities in reasoning, decision-making, and tool use~\cite{yao2022react,schick2023toolformer}. Recent work further enables agents to automatically construct workflows by selecting and orchestrating tools for a given task~\cite{zhang2024aflow,fernando2023promptbreeder,yuksekgonul2024textgrad,yang2023large, li2024autoflow}. However, existing workflow generation methods primarily optimize performance under benign conditions and rarely consider robustness to input corruptions or distribution shifts. This limitation is particularly critical for audio deepfake detection, where input signals are frequently altered by both system-level processes and environmental factors during transmission, storage, and platform processing.

This limitation is particularly important for audio deepfake detection, where different detectors exhibit varying robustness across acoustic conditions. Existing multi-detector pipelines typically rely on manually designed, fixed combinations of detectors and execution strategies, which may become suboptimal under unseen perturbations. This motivates a fundamental question: \textbf{\textit{Can LLM-based agents automatically construct robust audio deepfake detection workflows that generalize across diverse perturbed conditions?}}

In this paper, we introduce \textbf{ROGUE}: a \underline{R}obust w\underline{O}rkflow \underline{G}eneration framework via adversarial learning for a\underline{U}dio deepfake d\underline{E}tection. ROGUE formulates workflow generation as sequential decision-making under input perturbations and jointly optimizes a perturbation agent that generates challenging corruptions and a policy agent that adaptively orchestrates detection tools. This adversarial formulation encourages perturbation-aware tool selection and adaptive workflow construction, improving robustness and cross-dataset generalization. We evaluate ROGUE across multiple audio deepfake datasets and real-world corruptions, including environmental noise, compression, and neural codec distortions. Results show consistent improvements over static and non-adversarial workflow generation baselines.
\begin{itemize}[leftmargin=*]
    \item We propose ROGUE, a robust workflow generation framework for audio deepfake detection that formulates workflow construction as a sequential decision-making problem under corrupted perturbations, and introduces a dual-agent optimization paradigm consisting of a perturbation agent and a workflow policy agent.
    \item We demonstrate through extensive experiments that adversarially optimized workflow generation substantially improves robustness and cross-dataset generalization under diverse real-world corruptions.
    \item We provide detailed analyses of perturbation-aware workflow behavior, including perturbation importance, stage-wise robustness contributions, and workflow permutation studies, showing how adaptive multi-stage tool orchestration improves the robustness of audio deepfake detection.
\end{itemize}

\section{Related work}
\label{sec:related_work}

\textbf{LLM-based Workflow Generation.} Recent advances in large language models (LLMs) have enabled autonomous agents that plan and execute multi-step workflows. ReAct~\cite{yao2022react} and Toolformer~\cite{schick2023toolformer} integrate model 
reasoning and action, allowing LLMs to interleave chain-of-thought reasoning with environment/tool interactions, while HuggingGPT~\cite{shen2023hugginggpt} uses an LLM to coordinate task planning, model selection, execution, and response synthesis. Voyager~\cite{wang2023voyager} further enables agents to acquire and reuse skills through iterative prompting and feedback, maintaining a growing skill library for long-horizon tasks~\cite{wang2023voyager}. These works establish LLMs as planners and orchestrators of computational workflows.

More recent studies automate the design and optimization of such workflows. Prompt optimization methods~\cite{fernando2023promptbreeder, yang2023large, yuksekgonul2024textgrad, khattab2024dspy} represent LLM pipelines as composable programs and improve them through prompt compilation, bootstrapping, and feedback. Automated workflow optimization methods~\cite{khattab2024dspy, li2024autoflow, hu2024automated, zhang2024aflow, zhuge2024gptswarm} further optimize entire workflow structures, often through program synthesis or policy optimization. However, these methods generally assume benign inputs and optimize average-case performance without modeling how corrupted inputs affect workflow decisions. Recent work has examined workflow robustness under semantic variations and noisy instructions~\cite{xu2025robustflow}, but primarily from the perspective of instruction stability or reasoning consistency. In contrast, our framework jointly optimizes a perturbation agent and a workflow agent, casting workflow generation as robust decision-making under distribution shift.

\textbf{Audio Deepfake Detection.} Advances in TTS have enabled highly realistic synthetic speech, creating risks of misinformation, impersonation, and fraud. Early detectors use convolutional~\cite{tak2021endrawnet} and spectro-temporal architectures~\cite{jung2022aasist} to capture spoofing artifacts from spectrograms or raw waveforms. More recent methods leverage self-supervised speech foundation models, including Wav2Vec2.0 and HuBERT~\cite{wang2021investigating}, XLS-R~\cite{zhang2024audio}, and Wav2Vec-BERT~\cite{li2025we}, to obtain richer and more transferable representations. Despite these advances, detection performance often degrades under acoustic corruptions, codec distortions, unseen synthesis models, and adversarial perturbations~\cite{li2025measuring}. Existing robustness methods mainly improve individual detectors through data augmentation and perturbation-based training~\cite{zhang2024can, li2025measuring, tak2022rawboost}. In contrast, our work generates adaptive workflows that select and combine multiple detectors according to input characteristics.

\section{Method}
\label{sec:method}

\begin{figure*}[!t]
\centering
\includegraphics[width=0.7\textwidth]{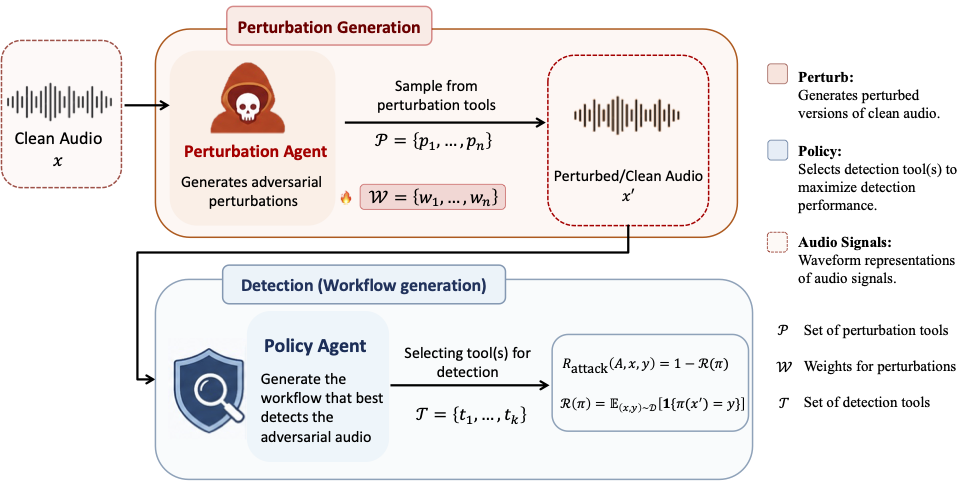}
\vspace{-0.2cm}
\caption{\small Overview of ROGUE. The framework consists of two interacting agents: a perturbation agent that generates audio perturbations by sampling from a set of perturbation tools, and a policy agent that dynamically constructs detection workflows by selecting and orchestrating detection tools. Through adversarial optimization, the workflow generator learns robust and adaptive detection strategies under perturbed conditions. 
}
\label{fig:overview}
\vspace{-0.5cm}
\end{figure*} 

We propose an adversarial learning framework for robust workflow generation in audio deepfake detection. Inspired by min–max adversarial learning~\cite{goodfellow2020generative,goodfellow2014explaining}, we extend robustness from individual models to workflow-level decision making. As shown in Figure~\ref{fig:overview}, our framework consists of (1) a \emph{perturbation agent} that generates perturbed audio and (2) a \emph{policy agent} that dynamically constructs detection workflows by selecting and orchestrating multiple detection tools.

\subsection{Perturbation Agent}

To simulate realistic and challenging input conditions, we introduce a perturbation agent that generates perturbed audio samples. 
We define a set of perturbation tools $\mathcal{P} = \{p_1, \dots, p_N\}$, where each perturbation $p_i$ corresponds to a realistic transformation commonly encountered in practice, e.g., pitch shifting, compression, or additive noise. Rather than applying perturbations uniformly, the agent maintains a learnable distribution over these transformations. At each step, the perturbation agent samples one perturbation type from the perturbation type set according to a categorical distribution parameterized by a weight vector: $ W = (w_1, \dots, w_N), \quad \text{where } \sum_{i=1}^N w_i = 1$, which determines how likely each perturbation is to be applied.
A perturbation is sampled as $p \sim \text{Categorical}(W)$, where $\text{Categorical}(W)$ denotes a discrete distribution over $\mathcal{P}$ defined by the probabilities in $W$. The perturbed audio is then generated as $x' = p(x)$. This design reflects practical scenarios where audio is typically affected by a dominant transformation (e.g., compression), while still allowing the agent to focus on the most challenging perturbations through adaptive reweighting of $W$.

The perturbation agent is trained adversarially to expose weaknesses in the detection workflow. Its objective is to generate perturbations that degrade the performance of the policy agent. We define its reward as:
\begin{equation}
R_{\text{perturbation}}(p, x, y) = 1 - R(\pi, x', y).
\end{equation}

where $R(\pi, x', y)$ denotes the reward function of the policy agent (defined in Equation~\ref{eq:reward} in Section~\ref{sec:policy_agent}). Intuitively, perturbations that lead to incorrect predictions or reduced confidence are assigned higher rewards, encouraging the perturbation agent to concentrate on perturbations that are more effective at degrading detection performance.

\subsection{Policy Agent (Workflow Generator)}
\label{sec:policy_agent}

The policy agent constructs a detection workflow by sequentially selecting and executing tools from a predefined set $\mathcal{T} = \{t_1, \dots, t_K\}$, which includes both lightweight heuristics and more computationally intensive deep neural detectors. Rather than relying on a fixed pipeline, the agent adaptively determines which tools to invoke and in what order based on intermediate observations. At step $i$, the agent maintains a state $ s_i = \left(x, \{o_1, \dots, o_{i-1}\}\right)$, where $x$ is the input audio and $\{o_1, \dots, o_{i-1}\}$ are the outputs of previously executed tools. Each observation $o_j$ may include prediction scores, confidence estimates, and auxiliary metadata, such as sample rate, durations, spectral artifact statistics, etc., providing contextual signals for subsequent decisions.

The policy is implemented as an LLM-based controller that maps the current state to an action: $\pi_\theta(a_i \mid s_i)$, where the action $a_i$ corresponds to either selecting the next tool or terminating the workflow. If a tool $t_{a_i}$ is selected, it is applied to the input $o_i = t_{a_i}(x)$, and the resulting observation is incorporated into the state. This iterative process allows the agent to refine its decisions based on accumulated evidence. Once the agent decides to terminate after $T$ steps, it produces a final prediction $\hat y = f_{\text{LLM}}(s_T)$, where $s_T$ is the final accumulated state. 
The overall procedure defines a sequential decision-making problem in which the LLM agent adaptively determines both tool selection and execution order based on intermediate observations.

We train the policy agent using a reinforcement learning formulation, where the environment consists of the perturbed input audio and the intermediate outputs produced by selected tools. The objective is to maximize the cumulative reward signal:

\begin{equation}
\label{eq:reward}
R(\pi) = \mathbb{E}_{(x,y) \sim \mathcal{D}} \left[ \mathbf{1}[\hat{y} = y] \right].
\end{equation}

To promote efficiency, we incorporate a cost penalty:
\begin{equation}
R(\pi) = \mathbb{E}_{(x,y)} \left[ \mathbf{1}[\hat{y} = y] - \lambda \cdot C(\pi) \right],
\end{equation}
where $C(\pi)$ measures computational cost, such as runtime or the number of tools used. This formulation encourages the policy to adopt adaptive escalation strategies, relying on inexpensive tools for simple cases and invoking more complex detectors only when necessary.

\subsection{Adversarial Learning Objective}

Unlike standard adversarial learning that improves the robustness of a fixed model, our framework jointly optimizes two interacting agents: a perturbation agent that generates challenging inputs and a policy agent that constructs robust workflows. This yields the min–max objective:
\begin{equation}
\pi^* = \arg\max_{\pi \in \Pi} \;
\mathbb{E}_{(x,y)\sim\mathcal{D}}
\left[
\min_{p \sim \text{Categorical}(W)}
R(\pi, p(x), y)
\right].
\end{equation}
The perturbation agent minimizes the policy’s reward, while the policy agent learns to adaptively select and combine tools under challenging perturbations. Thus, our approach enables robustness to emerge from adaptive workflow construction, allowing the system to remain effective under distribution shifts and diverse corruption patterns.

\section{Experiments}
\label{sec:exp}

\textbf{Datasets.} We train ROGUE on WaveFake~\cite{frank2021wavefake} and evaluate its robustness under diverse perturbations using the corresponding test split. Real audio samples are obtained from LJSpeech~\cite{ljspeech17}, with a balanced 1:1 real-to-fake ratio. To assess cross-dataset generalization under clean conditions, we conduct experiments on a diverse collection of benchmark datasets, including SONAR~\cite{li2025we}, ASVspoof2019~\cite{wang2020asvspoof}, ASVspoof2021 LA~\cite{yamagishi2021asvspoof}, CodecFake~\cite{xie2025codecfake}, Fake-or-Real~\cite{reimao2019dataset}, LibriSeVoc~\cite{Sun_2023_CVPR}, In-the-Wild~\cite{muller2022does}, and DFADD~\cite{du2024dfadd}.

\textbf{Evaluation Metrics.} We use classification accuracy as the primary evaluation metric. In practical deployment settings, audio deepfake detection systems are typically operated using fixed decision thresholds, making accuracy a more direct measure of real-world performance. Specifically, accuracy reflects the proportion of audio samples correctly classified as genuine or synthetic under a fixed operating point.

\textbf{Implementation Details.} We use GPT-5, Claude Sonnet 4.6, and Gemini 3 Pro as the backbone models for the workflow generator. Results using GPT-5 are presented in the main paper, while results for the remaining models are provided in the Appendix. Across all backbone models, we observe that ROGUE demonstrates consistent improvements in robustness and generalization compared to existing strong baselines. The policy agent is implemented as a DSPy-based workflow generator that receives an input audio sample together with intermediate detector outputs and produces a sequence of tool-selection actions forming a detection workflow. The DSPy program is optimized using a metric-driven objective that balances detection accuracy and computational cost, defined as the correctness of the final prediction minus a penalty proportional to the cumulative cost of executed tools. The perturbation agent operates over a predefined set of realistic audio perturbations, including pitch shifting, compression, and noise injection, following~\cite{li2025measuring}.

\textbf{Baselines.} We compare our framework against two self-supervised speech foundation models, Wav2Vec2-BERT and HuBERT from~\cite{li2025we}, which demonstrated strong detection performance. We also evaluate against DF\_Arena\_500M\_V1 and DF\_Arena\_1B\_V1~\cite{kulkarni2026compact}, which are currently ranked as the top two open-source detectors on the Speech DF Arena benchmark~\cite{11345101}. We also include a DSPy-optimized workflow without adversarial learning to isolate the contribution of adversarial optimization, providing a comprehensive comparison against existing approaches.

\begin{table}[t]
\centering

\resizebox{\columnwidth}{!}{
\begin{tabular}{lcccccccc}
\toprule
Method & Opus & EnCodec & Pitch & Time & Quant. & MP3 & HighPass & LowPass \\
\midrule
HuBERT & 0.64 & 0.56 & 0.71 & 0.73 & 0.68 & 0.97 & 0.96 & 0.98 \\
Wav2Vec2BERT & 0.69 & 0.59 & 0.67 & 0.75 & 0.71 & 0.98 & 0.99 & 0.99 \\
DF\_Arena\_1B\_V1 & 0.68 & 0.61 & 0.74 & 0.70 & 0.75 & 0.98 & 0.99 & 0.99 \\
DF\_Arena\_500M\_V1 & 0.66 & 0.55 & 0.73 & 0.68 & 0.74 & 0.98 & 0.99 & 0.99 \\
DSPy & 0.65 & 0.62 & 0.70 & 0.68 & 0.73 & 0.98 & 0.99 & 0.99 \\
\textbf{ROGUE} (ours) & \textbf{0.72} & \textbf{0.67} & \textbf{0.76} & \textbf{0.77} & \textbf{0.79} & 0.98 & 0.99 & 0.99 \\
\bottomrule
\end{tabular}
}

\vspace{0.1cm}

\resizebox{\columnwidth}{!}{
\begin{tabular}{lcccccccc}
\toprule
Method & Smooth & BN & BM & Echo & Autotune & Gauss. & Silent & Average \\
\midrule
HuBERT & 1.00 & 0.97 & 0.96 & 0.95 & 0.93 & 0.99 & 0.99 & 0.87 \\
Wav2Vec2BERT & 1.00 & 0.98 & 0.95 & 0.98 & 0.99 & 0.99 & 0.99 & 0.88 \\
DF\_Arena\_1B\_V1 & 1.00 & 0.98 & 0.97 & 0.98 & 0.98 & 0.99 & 0.99 & 0.89 \\
DF\_Arena\_500M\_V1 & 1.00 & 0.98 & 0.96 & 0.97 & 0.97 & 0.99 & 0.99 & 0.89 \\
DSPy & 1.00 & 0.97 & 0.97 & 0.97 & 0.96 & 0.99 & 0.99 & 0.88 \\
\textbf{ROGUE} (ours) & 1.00 & \textbf{0.99} & \textbf{0.97} & \textbf{0.98} & \textbf{0.99} & 0.99 & 0.99 & \textbf{0.91} \\
\bottomrule
\end{tabular}
}
\small
\caption{\small Robustness under diverse audio perturbations evaluated using the GPT-5-based workflow generation.}
\label{tab:perturbation}
\vspace{-0.2cm}
\end{table}

\begin{table}[t]
\centering

\resizebox{\columnwidth}{!}{
\begin{tabular}{ccccccccccc}
\toprule
Method & \begin{tabular}[c]{@{}c@{}}ASV\\2019\end{tabular} & \begin{tabular}[c]{@{}c@{}}Codec\\Fake\end{tabular} & \begin{tabular}[c]{@{}c@{}}ASV\\2021LA\end{tabular} & \begin{tabular}[c]{@{}c@{}}Fakeor\\Real\end{tabular} & DFADD & \begin{tabular}[c]{@{}c@{}}In-the \\-wild\end{tabular}  & SONAR & \begin{tabular}[c]{@{}c@{}}LibriSe\\Voc\end{tabular} & Average \\
\midrule
HuBERT & 0.89 & 0.71 & 0.82 & 0.90 & 1.00 & 0.91 & 0.87 & 0.99 & 0.89 \\
Wav2Vec2BERT & 0.93 & 0.74 & 0.88 & 0.86 & 1.00 & 0.92 & 0.89 & 0.98 & 0.90 \\
DF\_Arena\_1B\_V1 & 0.98 & 0.91 & 0.95 & 0.97 & 1.00 & 0.99 & 0.99 & 0.99 & 0.97 \\
DF\_Arena\_500M\_V1 & 0.98 & 0.93 & 0.96 & 0.97 & 1.00 & 0.98 & 0.98 & 1.00 & 0.97 \\
DSPy & 0.91 & 0.78 & 0.91 & 0.91 & 1.00 & 0.92 & 0.92 & 1.00 & 0.92 \\
\textbf{ROGUE} (ours) & 0.94 & 0.87 & 0.93 & 0.97 & 1.00 & 0.95 & 0.94 & 1.00 & 0.95 \\
\bottomrule
\end{tabular}}
\small
\caption{\small Cross-dataset generalization performance evaluated using the GPT-5-based workflow generation.}
\label{tab:generalization}
\vspace{-0.2cm}
\end{table}

\subsection{Robustness under Audio Perturbations}

We first evaluate the robustness of different methods under diverse realistic audio perturbations, including codec compression (Opus, EnCodec), signal transformations (pitch shift, time-stretching, smoothing), frequency filtering (high-pass and low-pass filtering), and environmental corruptions (background noise (BN), background music (BM), echo, and Gaussian noise). These perturbations simulate common degradations encountered in real-world audio transmission, recording, and post-processing pipelines.

\textbf{Table~\ref{tab:perturbation}} summarizes the robustness performance across all perturbation types. ROGUE consistently achieves the best average performance, outperforming both speech foundation model baselines and state-of-the-art specialized audio deepfake detectors. In particular, ROGUE achieves the largest gains under challenging perturbations that significantly alter the acoustic characteristics of the signal, including codec distortions (Opus and EnCodec), temporal modifications (Pitch and Time), and quantization artifacts. We also observe that codec-based perturbations remain among the most difficult corruptions for all methods. For example, HuBERT achieves only $0.56$ accuracy under EnCodec compression, whereas ROGUE improves performance to $0.67$. Similarly, under Opus compression, ROGUE achieves an accuracy of $0.72$, outperforming all competing baselines. These results suggest that adversarial workflow optimization enables the policy agent to adaptively select detectors that are more resilient to codec-induced distortions. Comparing DSPy-based workflow optimization with and without adversarial learning further demonstrates the importance of robustness-aware optimization. Although DSPy improves adaptive tool orchestration, incorporating adversarial learning leads to stronger robustness and yields more stable performance across diverse corruption types. These findings indicate that robustness in workflow generation cannot be achieved solely by optimizing for clean or benign inputs, and explicitly modeling against perturbations is critical for reliable deployment in real-world environments.

\subsection{Cross-Dataset Generalization}

We then evaluate the cross-dataset generalization capability of different methods across a diverse collection of benchmark datasets. The results are summarized in \textbf{Table~\ref{tab:generalization}}. Overall, ROGUE achieves consistently strong performance across nearly all evaluation datasets, demonstrating improved robustness and generalization compared to both single-model baselines and non-adversarial workflow optimization approaches. In particular, ROGUE achieves notable gains on challenging out-of-distribution benchmarks such as CodecFake and In-the-Wild, improving performance by approximately $9\%$ and $3\%$ over the DSPy baseline, respectively. These datasets are especially difficult because they contain substantial distribution shifts caused by neural codec compression, real-world recording artifacts, and unseen synthesis conditions. 
We also observe that DSPy-based workflow optimization consistently improves performance over standalone foundation models, suggesting that adaptive tool orchestration provides benefits beyond relying on individual detectors alone. However, DSPy optimization without adversarial learning still underperforms compared to ROGUE. The results indicate that optimizing workflows solely on clean or benign inputs is insufficient for achieving strong cross-dataset generalization, and that robustness-aware adversarial optimization can further improve generalization under distribution shifts.

\subsection{Perturbation Analysis}

\begin{figure}
    \centering
    \includegraphics[width=0.8\columnwidth]{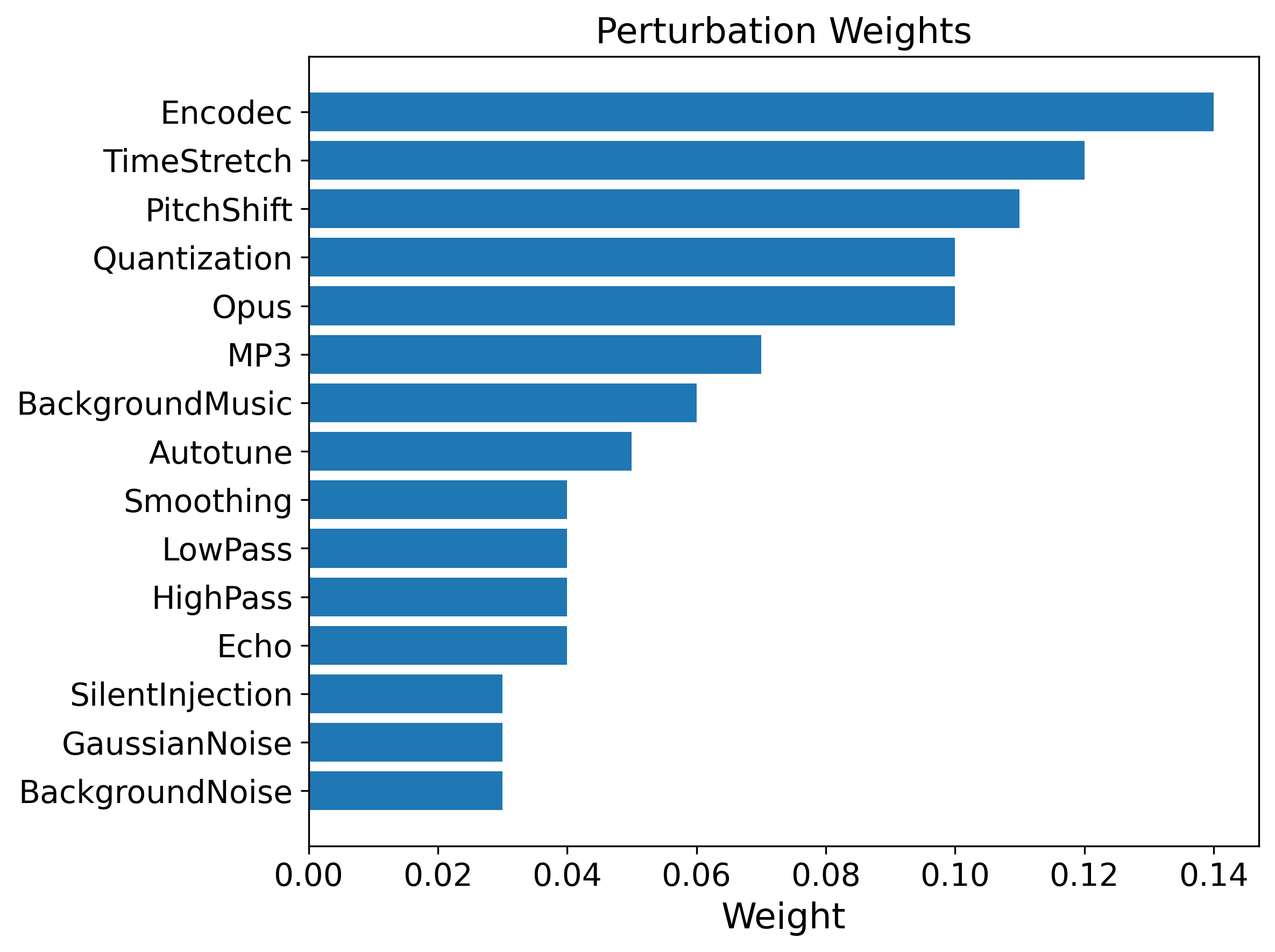}
    \caption{\small Perturbation weights.}
    \label{fig:weights}
    \vspace{-0.4cm}
\end{figure}

\textbf{Figure~\ref{fig:weights}} illustrates the perturbation weights learned by the perturbation agent. Consistent with prior robustness studies~\cite{li2025measuring}, challenging perturbations such as EnCodec, Opus, quantization, pitch shifting, and time stretching receive high weights, whereas Gaussian noise and silence injection receive much lower weights.

To further examine the contribution of individual perturbation types, we further conduct a leave-one-out study by removing each perturbation from adversarial learning while keeping all other settings fixed. As shown in \textbf{Table~\ref{tab:perturb_leave_one_out}}, removing EnCodec causes the largest drop in average robust accuracy, from 0.74 to 0.68. Excluding Opus or pitch perturbations also noticeably reduces robustness, while removing quantization or time stretching has a smaller effect. These results show that the perturbation agent effectively prioritizes transformations that most strongly affect workflow robustness.

\begin{table}

    \resizebox{0.8\columnwidth}{!}{
    \begin{tabular}{lcc}
    \toprule
    Training Setup & Avg. Robust Acc. & Drop \\
    \midrule
    ROGUE (All) & 0.74 & 0.00 \\
    w/o Opus    & 0.71 & -0.03 \\
    w/o EnCodec & 0.68 & -0.06 \\
    w/o Pitch   & 0.70 & -0.04 \\
    w/o Time    & 0.73 & -0.01 \\
    w/o Quant.  & 0.74 & -0.00 \\
    \bottomrule
    \end{tabular}}
\centering
\small
    \captionof{table}{\small Leave-one-out perturbation analysis. Larger performance drops indicate more effective perturbations during adversarial workflow training.}
    \label{tab:perturb_leave_one_out}
\vspace{-0.4cm}
\end{table}

\begin{figure*}[!ht]
\centering
\includegraphics[width=0.7\textwidth]{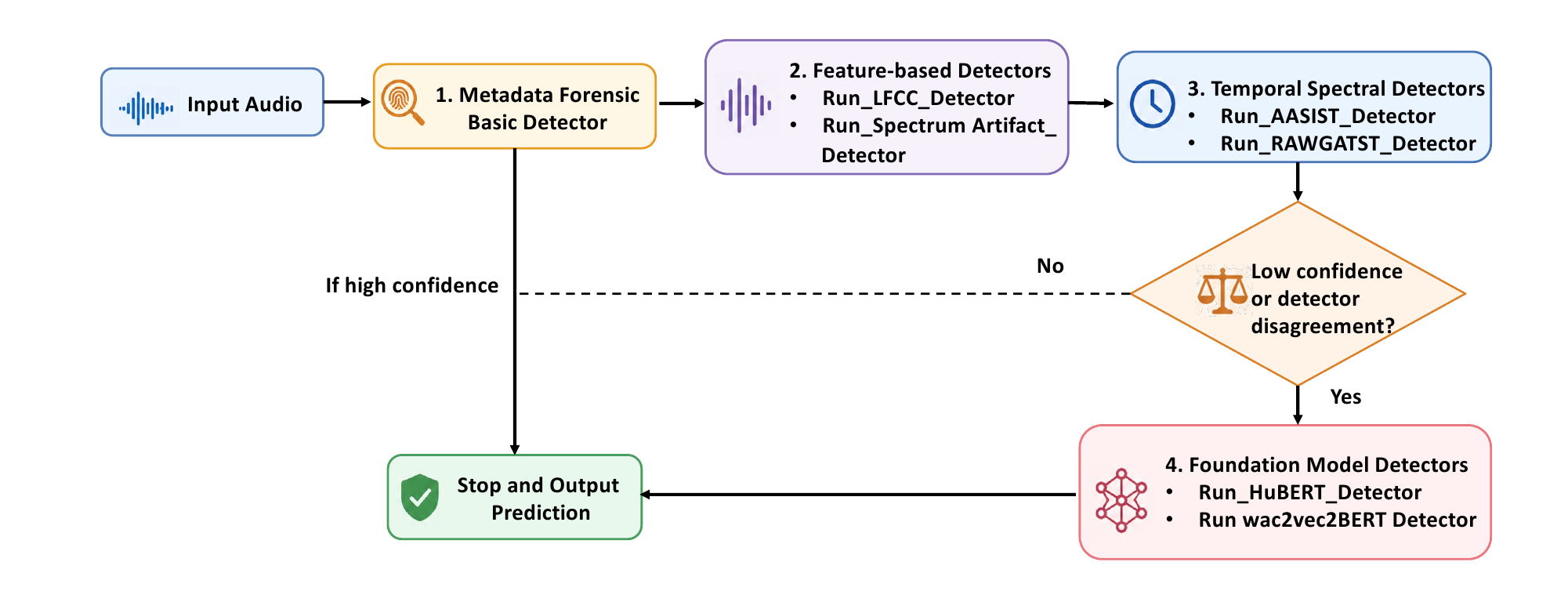}

\caption{\small Workflow generated by ROGUE.}
\label{fig:workflow}

\end{figure*} 

\subsection{Workflow analysis}

\textbf{Figure~\ref{fig:workflow}} in the Appendix illustrates the optimized detection workflow. To better understand the contribution of each stage, we conduct a stage-wise robustness analysis on the WaveFake test set under all perturbation settings. The analysis is performed from two complementary perspectives: (1) progressively enabling additional stages in the workflow pipeline (Stage 1, Stage 1+2, Stage 1+2+3, and Stage 1+2+3+4) to evaluate how robustness evolves as more advanced detectors are incorporated; and (2) independently evaluating each stage using only the detectors within that stage to measure the standalone robustness of different detector categories. Results are reported as the average detection accuracy across all perturbation types. 
\begin{table}

\resizebox{0.88\columnwidth}{!}{
\begin{tabular}{cc}
\toprule
Workflow Configuration & Avg. Robust Accuracy \\
\midrule
Stage 1 & 0.68 \\
Stage 1 + 2 & 0.72 \\
Stage 1 + 2 + 3 & 0.76 \\
Stage 1 + 2 + 3 + 4 & \textbf{0.95} \\
\bottomrule
\end{tabular}}
\centering
\small
\caption{\small Average robustness accuracy across all perturbations using progressively accumulated workflow stages.}
\label{tab:workflow_accumulation}
\vspace{-0.2cm}
\end{table}

\textbf{Table~\ref{tab:workflow_accumulation}} presents the accumulated workflow results. Robustness consistently improves as additional stages are incorporated into the pipeline. The early stages, which primarily consist of lightweight detectors, provide limited robustness due to their reliance on shallow acoustic and metadata cues. Adding feature-based detectors improves robustness by capturing spectral artifacts associated with synthetic audio generation, while temporal-spectral analysis further enhances robustness performance by modeling temporal inconsistencies in deepfake audios. The largest improvement occurs when foundation-model-based detectors are introduced, highlighting the importance of high-capacity semantic representations under challenging perturbation conditions.

\begin{table}[t]
\centering
\small
\resizebox{0.7\columnwidth}{!}{
\begin{tabular}{lc}
\toprule
Stage & Avg. Robust Accuracy \\
\midrule
Stage 1 & 0.68 \\
Stage 2 & 0.74 \\
Stage 3 & 0.75 \\
Stage 4 & \textbf{0.92} \\
\bottomrule
\end{tabular}}

\caption{\small Average robustness accuracy across all perturbations using only detectors from each individual stage.}
\label{tab:individual_stage}
\vspace{-0.4cm}
\end{table}

\textbf{Table~\ref{tab:individual_stage}} reports the standalone evaluation of each stage. Later stages achieve substantially higher robustness than earlier lightweight detectors, with Stage 4 producing the strongest standalone performance. This result indicates that foundation-model-based detectors contribute the majority of robustness gains. However, the progressively accumulated workflow still outperforms any individual stage alone, suggesting that the multi-stage design provides complementary information and enables efficient early exits for high-confidence samples while reserving computationally expensive detectors for more difficult cases.

\begin{table}
\centering
\small

\resizebox{0.6\columnwidth}{!}{
\begin{tabular}{lcc}
\toprule
Workflow Order & Robust Accuracy \\
\midrule
1 $\rightarrow$ 2 $\rightarrow$ 3 $\rightarrow$ 4 & \textbf{0.95}  \\
1 $\rightarrow$ 3 $\rightarrow$ 2 $\rightarrow$ 4 & 0.90  \\
2 $\rightarrow$ 1 $\rightarrow$ 3 $\rightarrow$ 4 & 0.88  \\
3 $\rightarrow$ 2 $\rightarrow$ 1 $\rightarrow$ 4 & 0.84  \\
2 $\rightarrow$ 4 $\rightarrow$ 1 $\rightarrow$ 3 & 0.87  \\
4 $\rightarrow$ 1 $\rightarrow$ 2 $\rightarrow$ 3 & 0.92  \\
\bottomrule
\end{tabular}}
\caption{\small Representative workflow permutation results.}
\label{tab:workflow_permutation}
\vspace{-0.5cm}
\end{table}

To further evaluate the soundness of the optimized workflow structure, we conduct a workflow permutation analysis by testing alternative stage orderings of the proposed pipeline. \textbf{Table~\ref{tab:workflow_permutation}} reports representative permutations and their average robustness accuracy across all perturbation types. The proposed ordering achieves the best performance, while alternative permutations lead to robustness degradation. In particular, placing lightweight detectors in earlier stages and progressively escalating difficult samples to more sophisticated detectors yields the most effective coarse-to-fine detection strategy. These results suggest that the workflow structure itself plays an important role beyond simply combining multiple detectors.

\section{Discussion}
\label{sec:discussion}

Despite its advantages, ROGUE has several limitations: (1) Search complexity: the joint search space over perturbation compositions and workflow sequences grows exponentially, making optimization computationally expensive. (2) Dependence on tool diversity: the effectiveness of the framework depends on the diversity and availability of complementary detection tools; limited tool diversity may restrict performance gains.
(3) Although the perturbation agent models a broad range of realistic audio perturbations, the perturbation agent may not exhaustively capture real-world distortions or adversarial manipulation strategies encountered in deployment environments. (4) While ROGUE achieves substantially improved robustness, it incurs a modest increase in computational cost compared to single-model baselines. However, this trade-off is explicit and controllable: the learned workflows dynamically decide when additional computation is necessary. In practice, such adaptive resource allocation is often more desirable than fixed-cost detection pipelines.

\vspace{-0.3cm}
\section{Conclusion}
\label{sec:conclusion}
\vspace{-0.3cm}

In this paper, we introduced ROGUE, an adversarial learning framework for robust workflow generation in audio deepfake detection. ROGUE jointly optimizes a perturbation agent that generates challenging audio corruptions and a policy agent that dynamically selects and orchestrates detection tools. Extensive experiments demonstrate improved robustness and cross-dataset generalization under diverse perturbations. Although ROGUE was initially designed for audio deepfake detection, ROGUE offers a general paradigm for constructing robust agentic workflows, with potential applications to LLM-based systems operating under distribution shifts, noisy inputs, and corrupted environments.

\bibliographystyle{IEEEbib}
\bibliography{main.bbl}

\begin{thebibliography}{10}

\bibitem{vyas2023audiobox}
Apoorv Vyas, Bowen Shi, Matthew Le, Andros Tjandra, Yi-Chiao Wu, Baishan Guo, Jiemin Zhang, Xinyue Zhang, Robert Adkins, William Ngan, et~al.,
\newblock ``Audiobox: Unified audio generation with natural language prompts,''
\newblock {\em arXiv preprint arXiv:2312.15821}, 2023.

\bibitem{ye2024flashspeech}
Zhen Ye, Zeqian Ju, Haohe Liu, Xu~Tan, Jianyi Chen, Yiwen Lu, Peiwen Sun, Jiahao Pan, Weizhen Bian, Shulin He, et~al.,
\newblock ``Flashspeech: Efficient zero-shot speech synthesis,''
\newblock {\em arXiv preprint arXiv:2404.14700}, 2024.

\bibitem{wang2021investigating}
Xin Wang and Junichi Yamagishi,
\newblock ``Investigating self-supervised front ends for speech spoofing countermeasures,''
\newblock {\em arXiv preprint arXiv:2111.07725}, 2021.

\bibitem{zhang2024audio}
Qishan Zhang, Shuangbing Wen, and Tao Hu,
\newblock ``Audio deepfake detection with self-supervised xls-r and sls classifier,''
\newblock in {\em Proceedings of the 32nd ACM International Conference on Multimedia}, 2024, pp. 6765--6773.

\bibitem{li2025we}
Xiang Li, Pin-Yu Chen, and Wenqi Wei,
\newblock ``Where are we in audio deepfake detection? a systematic analysis over generative and detection models,''
\newblock {\em ACM Transactions on Internet Technology}, vol. 25, no. 3, pp. 1--19, 2025.

\bibitem{li2025measuring}
Xiang Li, Pin-Yu Chen, and Wenqi Wei,
\newblock ``Measuring the robustness of audio deepfake detection under real-world corruption,''
\newblock in {\em Proceedings of the Sixteenth ACM Conference on Data and Application Security and Privacy}, 2026.

\bibitem{yao2022react}
Shunyu Yao, Jeffrey Zhao, Dian Yu, Nan Du, Izhak Shafran, Karthik Narasimhan, and Yuan Cao,
\newblock ``React: Synergizing reasoning and acting in language models,''
\newblock {\em arXiv preprint arXiv:2210.03629}, 2022.

\bibitem{schick2023toolformer}
Timo Schick, Jane Dwivedi-Yu, Roberto Dess{\`\i}, Roberta Raileanu, Maria Lomeli, Eric Hambro, Luke Zettlemoyer, Nicola Cancedda, and Thomas Scialom,
\newblock ``Toolformer: Language models can teach themselves to use tools,''
\newblock {\em Advances in neural information processing systems}, vol. 36, pp. 68539--68551, 2023.

\bibitem{zhang2024aflow}
Jiayi Zhang, Jinyu Xiang, Zhaoyang Yu, Fengwei Teng, Xionghui Chen, Jiaqi Chen, Mingchen Zhuge, Xin Cheng, Sirui Hong, Jinlin Wang, et~al.,
\newblock ``Aflow: Automating agentic workflow generation,''
\newblock {\em arXiv preprint arXiv:2410.10762}, 2024.

\bibitem{fernando2023promptbreeder}
Chrisantha Fernando, Dylan Banarse, Henryk Michalewski, Simon Osindero, and Tim Rockt{\"a}schel,
\newblock ``Promptbreeder: Self-referential self-improvement via prompt evolution,''
\newblock {\em arXiv preprint arXiv:2309.16797}, 2023.

\bibitem{yuksekgonul2024textgrad}
Mert Yuksekgonul, Federico Bianchi, Joseph Boen, Sheng Liu, Zhi Huang, Carlos Guestrin, and James Zou,
\newblock ``Textgrad: Automatic" differentiation" via text,''
\newblock {\em arXiv preprint arXiv:2406.07496}, 2024.

\bibitem{yang2023large}
Chengrun Yang, Xuezhi Wang, Yifeng Lu, Hanxiao Liu, Quoc~V Le, Denny Zhou, and Xinyun Chen,
\newblock ``Large language models as optimizers,''
\newblock in {\em The Twelfth International Conference on Learning Representations}, 2023.

\bibitem{li2024autoflow}
Zelong Li, Shuyuan Xu, Kai Mei, Wenyue Hua, Balaji Rama, Om~Raheja, Hao Wang, He~Zhu, and Yongfeng Zhang,
\newblock ``Autoflow: Automated workflow generation for large language model agents,''
\newblock {\em arXiv preprint arXiv:2407.12821}, 2024.

\bibitem{shen2023hugginggpt}
Yongliang Shen, Kaitao Song, Xu~Tan, Dongsheng Li, Weiming Lu, and Yueting Zhuang,
\newblock ``Hugginggpt: Solving ai tasks with chatgpt and its friends in hugging face,''
\newblock {\em Advances in Neural Information Processing Systems}, vol. 36, pp. 38154--38180, 2023.

\bibitem{wang2023voyager}
Guanzhi Wang, Yuqi Xie, Yunfan Jiang, Ajay Mandlekar, Chaowei Xiao, Yuke Zhu, Linxi Fan, and Anima Anandkumar,
\newblock ``Voyager: An open-ended embodied agent with large language models, 2023,''
\newblock {\em URL https://arxiv. org/abs/2305.16291}, vol. 2, no. 11, 2023.

\bibitem{khattab2024dspy}
Omar Khattab, Arnav Singhvi, Paridhi Maheshwari, Zhiyuan Zhang, Keshav Santhanam, Sri Vardhamanan, Saiful Haq, Ashutosh Sharma, Thomas~T. Joshi, Hanna Moazam, Heather Miller, Matei Zaharia, and Christopher Potts,
\newblock ``Dspy: Compiling declarative language model calls into self-improving pipelines,''
\newblock 2024.

\bibitem{hu2024automated}
Shengran Hu, Cong Lu, and Jeff Clune,
\newblock ``Automated design of agentic systems,''
\newblock {\em arXiv preprint arXiv:2408.08435}, 2024.

\bibitem{zhuge2024gptswarm}
Mingchen Zhuge, Wenyi Wang, Louis Kirsch, Francesco Faccio, Dmitrii Khizbullin, and J{\"u}rgen Schmidhuber,
\newblock ``Gptswarm: Language agents as optimizable graphs,''
\newblock in {\em Forty-first International Conference on Machine Learning}, 2024.

\bibitem{xu2025robustflow}
Shengxiang Xu, Jiayi Zhang, Shimin Di, Yuyu Luo, Liang Yao, Hanmo Liu, Jia Zhu, Fan Liu, and Min-Ling Zhang,
\newblock ``Robustflow: Towards robust agentic workflow generation,''
\newblock {\em arXiv preprint arXiv:2509.21834}, 2025.

\bibitem{tak2021endrawnet}
Hemlata Tak, Jose Patino, Massimiliano Todisco, Andreas Nautsch, Nicholas Evans, and Anthony Larcher,
\newblock ``End-to-end anti-spoofing with rawnet2,''
\newblock in {\em IEEE International Conference on Acoustics, Speech and Signal Processing (ICASSP)}, 2021, pp. 6369--6373.

\bibitem{jung2022aasist}
Jee-weon Jung, Hee-Soo Heo, Hemlata Tak, Hye-jin Shim, Joon~Son Chung, Bong-Jin Lee, Ha-Jin Yu, and Nicholas Evans,
\newblock ``Aasist: Audio anti-spoofing using integrated spectro-temporal graph attention networks,''
\newblock in {\em IEEE international conference on acoustics, speech and signal processing (ICASSP)}, 2022, pp. 6367--6371.

\bibitem{zhang2024can}
Zirui Zhang, Wei Hao, Aroon Sankoh, William Lin, Emanuel Mendiola-Ortiz, Junfeng Yang, and Chengzhi Mao,
\newblock ``I can hear you: Selective robust training for deepfake audio detection,''
\newblock {\em arXiv preprint arXiv:2411.00121}, 2024.

\bibitem{tak2022rawboost}
Hemlata Tak, Madhu Kamble, Jose Patino, Massimiliano Todisco, and Nicholas Evans,
\newblock ``Rawboost: A raw data boosting and augmentation method applied to automatic speaker verification anti-spoofing,''
\newblock in {\em ICASSP 2022-2022 IEEE International Conference on Acoustics, Speech and Signal Processing (ICASSP)}. IEEE, 2022, pp. 6382--6386.

\bibitem{goodfellow2020generative}
Ian Goodfellow, Jean Pouget-Abadie, Mehdi Mirza, Bing Xu, David Warde-Farley, Sherjil Ozair, Aaron Courville, and Yoshua Bengio,
\newblock ``Generative adversarial networks,''
\newblock {\em Communications of the ACM}, vol. 63, no. 11, pp. 139--144, 2020.

\bibitem{goodfellow2014explaining}
Ian~J Goodfellow, Jonathon Shlens, and Christian Szegedy,
\newblock ``Explaining and harnessing adversarial examples,''
\newblock {\em arXiv preprint arXiv:1412.6572}, 2014.

\bibitem{frank2021wavefake}
Joel Frank and Lea Sch{\"o}nherr,
\newblock ``Wavefake: A data set to facilitate audio deepfake detection,''
\newblock {\em arXiv preprint arXiv:2111.02813}, 2021.

\bibitem{ljspeech17}
Keith Ito and Linda Johnson,
\newblock ``The lj speech dataset,'' \url{https://keithito.com/LJ-Speech-Dataset/}, 2017.

\bibitem{wang2020asvspoof}
Xin Wang, Junichi Yamagishi, Massimiliano Todisco, H{\'e}ctor Delgado, Andreas Nautsch, Nicholas Evans, Md~Sahidullah, Ville Vestman, Tomi Kinnunen, Kong~Aik Lee, et~al.,
\newblock ``Asvspoof 2019: A large-scale public database of synthesized, converted and replayed speech,''
\newblock {\em Computer Speech \& Language}, vol. 64, pp. 101114, 2020.

\bibitem{yamagishi2021asvspoof}
Junichi Yamagishi, Xin Wang, Massimiliano Todisco, Md~Sahidullah, Jose Patino, Andreas Nautsch, Xuechen Liu, Kong~Aik Lee, Tomi Kinnunen, Nicholas Evans, et~al.,
\newblock ``Asvspoof 2021: accelerating progress in spoofed and deepfake speech detection,''
\newblock {\em arXiv preprint arXiv:2109.00537}, 2021.

\bibitem{xie2025codecfake}
Yuankun Xie, Yi~Lu, Ruibo Fu, Zhengqi Wen, Zhiyong Wang, Jianhua Tao, Xin Qi, Xiaopeng Wang, Yukun Liu, Haonan Cheng, et~al.,
\newblock ``The codecfake dataset and countermeasures for the universally detection of deepfake audio,''
\newblock {\em IEEE Transactions on Audio, Speech and Language Processing}, 2025.

\bibitem{reimao2019dataset}
Ricardo Reimao and Vassilios Tzerpos,
\newblock ``For: A dataset for synthetic speech detection,''
\newblock in {\em 2019 International Conference on Speech Technology and Human-Computer Dialogue (SpeD)}. IEEE, 2019, pp. 1--10.

\bibitem{Sun_2023_CVPR}
Chengzhe Sun, Shan Jia, Shuwei Hou, and Siwei Lyu,
\newblock ``Ai-synthesized voice detection using neural vocoder artifacts,''
\newblock in {\em Proceedings of the IEEE/CVF Conference on Computer Vision and Pattern Recognition (CVPR) Workshops}, June 2023, pp. 904--912.

\bibitem{muller2022does}
Nicolas~M M{\"u}ller, Pavel Czempin, Franziska Dieckmann, Adam Froghyar, and Konstantin B{\"o}ttinger,
\newblock ``Does audio deepfake detection generalize?,''
\newblock {\em Interspeech}, 2022.

\bibitem{du2024dfadd}
Jiawei Du, I-Ming Lin, I-Hsiang Chiu, Xuanjun Chen, Haibin Wu, Wenze Ren, Yu~Tsao, Hung-yi Lee, and Jyh-Shing~Roger Jang,
\newblock ``Dfadd: The diffusion and flow-matching based audio deepfake dataset,''
\newblock in {\em 2024 IEEE Spoken Language Technology Workshop (SLT)}. IEEE, 2024, pp. 921--928.

\bibitem{kulkarni2026compact}
Ajinkya Kulkarni, Sandipana Dowerah, Atharva Kulkarni, Tanel Alum{\"a}e, and Mathew~Magimai Doss,
\newblock ``Do compact ssl backbones matter for audio deepfake detection? a controlled study with raptor,''
\newblock {\em arXiv preprint arXiv:2603.06164}, 2026.

\bibitem{11345101}
Sandipana Dowerah, Atharva Kulkarni, Ajinkya Kulkarni, Hoan~My Tran, Joonas Kalda, Artem Fedorchenko, Benoit Fauve, Damien Lolive, Tanel Alumäe, and Mathew Magimai.-Doss,
\newblock ``Speech df arena: A leaderboard for speech deepfake detection models,''
\newblock {\em IEEE Open Journal of Signal Processing}, pp. 1--9, 2026.

\end{thebibliography}

\appendix

\section{Workflow Optimization Algorithm}

The overall adversarial learning procedure is summarized in Algorithm~\ref{alg:adv_workflow}. 

\begin{algorithm*}[t]
\small

\caption{\small Adversarial Learning for Robust Workflow Generation}
\label{alg:adv_workflow}
\begin{algorithmic}[1]
\Require Training data $\mathcal{D}$, perturbation set $\mathcal{P}=\{p_1,\dots,p_N\}$, detection tools $\mathcal{T}=\{t_1,\dots,t_K\}$, workflow policy $\pi_\theta$, perturbation weights $W$, epochs $E$, search steps $M$
\Ensure Robust workflow policy $\pi_\theta$

\State Initialize workflow policy $\pi_\theta$ with few-shot demonstrations $\mathcal{S}$
\State Initialize perturbation weights $W \gets \frac{1}{N}\mathbf{1}$

\For{$e = 1$ to $E$}
    \For{each mini-batch $\mathcal{B} \subset \mathcal{D}$}

        \State Initialize adversarial batch $\mathcal{B}_{adv} \gets \emptyset$, perturbation statistics $\mathcal{A} \gets \mathbf{0}$

        \For{each $(x,y) \in \mathcal{B}$}

            \State Initialize worst-case reward $r^* \gets +\infty$, perturbation $p^* \gets \emptyset$

            \For{$m = 1$ to $M$}

                \State Sample perturbation and generate perturbed audio $x'_m \gets p_m(x)$
                \State Initialize workflow state $s_0 \gets (x'_m)$

                \For{$i = 1$ to $T$}
                    \State Sample action $a_i \sim \pi_\theta(a_i \mid s_{i-1})$
                    \If{$a_i = \texttt{STOP}$}
                        \State break
                    \EndIf
                    \State Execute tool $o_i \gets t_{a_i}(x'_m)$ and update state $s_i \gets (s_{i-1}, o_i)$
                \EndFor

                \State Generate final prediction $\hat y_m \gets f_{\text{LLM}}(s_T)$
                \State Compute reward $r_m = \mathbf{1}[\hat y_m = y] - \lambda C(\pi_m)$

                \If{$r_m < r^*$}
                    \State $r^* \gets r_m$, $p^* \gets p_m$
                \EndIf

            \EndFor

            \State Generate adversarial example $x^* \gets p^*(x)$ and add $(x^*, y)$ to $\mathcal{B}_{adv}$
            \State Update perturbation statistic $\mathcal{A}_{p^*} \gets \mathcal{A}_{p^*} + (1-r^*)$

        \EndFor

        \State Update perturbation weights:
        \[
        W_i \gets \frac{\exp(\mathcal{A}_i/\tau)}
        {\sum_{j=1}^{N}\exp(\mathcal{A}_j/\tau)}
        \]

        \State Generate workflows for $\mathcal{B}_{adv}$ using $\pi_\theta$
        \State Compute batch reward:
        \[
        R(\pi_\theta) =
        \mathbb{E}_{(x^*,y)\in\mathcal{B}_{adv}}
        \left[
        \mathbf{1}[\hat y = y] - \lambda C(\pi_\theta)
        \right]
        \]

        \State Update workflow policy $\pi_\theta$ to maximize $R(\pi_\theta)$

    \EndFor
\EndFor

\State \Return $\pi_\theta$
\end{algorithmic}
\end{algorithm*}
\begin{table*}[t]
\centering

\resizebox{0.8\textwidth}{!}{
\begin{tabular}{lcccccccc}
\toprule
Method & Opus & EnCodec & Pitch & Time & Quant. & MP3 & HighPass & LowPass \\
\midrule
GPT-5 & 0.72 & 0.67 & 0.76 & 0.77 & 0.79 & 0.98 & 0.99 & 0.99 \\
Claude-sonnet-4-6 & 0.69 & 0.66 & 0.78 & 0.73 & 0.77 & 0.99 & 0.99 & 0.99 \\
Gemini-3-pro & 0.72 & 0.63 & 0.74 & 0.76 & 0.73 & 0.98 & 0.99 & 0.99 \\
\bottomrule
\end{tabular}
}

\vspace{0.1cm}

\resizebox{0.8\textwidth}{!}{
\begin{tabular}{lcccccccc}
\toprule
Method & Smoothing & BN & BM & Echo & Autotune & Gauss & Silent & Average \\
\midrule
GPT-5 & 1.00 & 0.99 & 0.97 & 0.98 & 0.99 & 0.99 & 0.99 & 0.91 \\
Claude-sonnet-4-6 & 1.00 & 0.99 & 0.97 & 0.98 & 0.99 & 0.99 & 1.00 & 0.90 \\
Gemini-3-pro & 1.00 & 0.99 & 0.98 & 0.98 & 0.99 & 1.00 & 0.99 & 0.90\\
\bottomrule
\end{tabular}
}
\caption{Robustness results under different perturbation types for different LLM-based workflow generation frameworks.}
\label{tab:appendix_perturbation}
\end{table*}

\begin{table*}[t]
\centering

\resizebox{0.8\textwidth}{!}{
\begin{tabular}{lccccccccc}
\toprule
Method & ASV2019 & CodecFake & ASV2021LA & FakeorReal & DFADD & In-the-Wild & SONAR & LibriSeVoc & Average \\
\midrule
GPT-5 & 0.942 & 0.878 & 0.937 & 0.974 & 0.998 & 0.953 & 0.942 & 1.000 &0.953 \\
Claude-sonnet-4-6 & 0.934 & 0.836 & 0.926 & 0.927 & 1.000 & 0.961 & 0.918 & 1.000 &0.938\\
Gemini-3-pro & 0.931 & 0.822 & 0.934 & 0.931 & 1.000 & 0.953 & 0.926 & 1.000&0.937 \\
\bottomrule
\end{tabular}
}
\caption{Cross-dataset robustness results for different LLM-based workflow generation frameworks.}
\label{tab:appendix_cross_dataset}
\end{table*}

\section{Detailed Implementation Details}

\textbf{Detection Tools.} For metadata forensics, we extract audio metadata and low-level acoustic statistics, including sample rate, approximate bitrate, and spectrogram-based features such as spectral centroid and spectral flatness. For audio deepfake detectors, we adopt CQCC-GMM as the basic lightweight detector. For feature-based detectors, we incorporate LFCC-LCNN, which leverages linear frequency cepstral coefficients (LFCC) to capture discriminative spectral characteristics of synthetic speech, and Spectrogram-ResNet, which utilizes spectrogram representations to detect visual spectral artifacts introduced during audio generation. For temporal-spectral detectors, we include AASIST and RawGAT-ST, which jointly model temporal and spectral dependencies to capture inconsistencies in synthetic audio signals. For foundation-model-based detectors, we incorporate HuBERT and Wav2Vec2-BERT, which are pretrained on massive-scale speech corpora and provide rich and transferable speech representations that demonstrate stronger generalization and robustness under distribution shifts and perturbations.

\textbf{Perturbation Tools.} For perturbation generation, we adopt 15 different corruption types spanning noise perturbations, audio modifications, and compression perturbations. Noise perturbations include background music, background noise, and Gaussian noise. Audio modification perturbations include pitch shifting, time stretching, high-pass filtering, low-pass filtering, smoothing, echo, autotune, and silence injection. Compression perturbations include Opus compression, MP3 compression, EnCodec neural codec compression, and quantization.

\section{Additional Results}
Tables~\ref{tab:appendix_perturbation} and~\ref{tab:appendix_cross_dataset} present results on robustness against different perturbations and cross-dataset generalization, respectively, using different LLM backbones as the policy generator. In particular, despite variations in performance across individual conditions, all three backbones achieve comparable overall performance, suggesting that our framework's effectiveness does not depend on a specific LLM backbone.

\end{document}